# ARPES Study of the Evolution of Band Structure and Charge Density Wave Properties in $RTe_3$ for R = Y, La, Ce, Sm, Gd, Tb and Dy

V. Brouet[1,2,3], W.L. Yang[2,3], X.J. Zhou[2,3], Z. Hussain[2], R.G. Moore[3,4], R. He[3,4], D.H. Lu[3], Z.X. Shen[2,3,4], J. Laverock[5], S. Dugdale[5], N. Ru[4] and I.R. Fisher[4]

[1]Lab. Physique des Solides, UMR8502, CNRS, Université Paris-Sud XI, Bât 510, 91405 Orsay (France)
[2]Advanced Light Source, Lawrence Berkeley National Laboratory, Berkeley, California 94720, USA
[3]Stanford Synchrotron Radiation Laboratory, Stanford University, Stanford, California 94305, USA
[4]Geballe Laboratory for Advanced Materials and Department of Applied Physics, Stanford University, Stanford, California 94305-4045, USA
[5]H. H. Wills Physics Laboratory, University of Bristol, Tyndall Avenue, Bristol BS8 1TL, United Kingdom

We present a detailed ARPES investigation of the $RTe_3$ family, which sets this system as an ideal "textbook" example for the formation of a nesting driven Charge Density Wave (CDW). This family indeed exhibits the full range of phenomena that can be associated to CDW instabilities, from the opening of large gaps on the best nested parts of Fermi Surface (FS) (up to 0.4eV), to the existence of residual metallic pockets. ARPES is the best suited technique to characterize these features, thanks to its unique ability to resolve the electronic structure in k-space. An additional advantage of $RTe_3$ is that the band structure can be very accurately described by a simple 2D tight-binding (TB) model, which allows one to understand and easily reproduce many characteristics of the CDW. In this paper, we first establish the main features of the electronic structure, by comparing our ARPES measurements with Linear Muffin-Tin Orbital band calculations. We use this to define the validity and limits of the TB model. We then present a complete description of the CDW properties and, for the first time, of their strong evolution as a function of R. Using simple models, we are able to reproduce perfectly the evolution of gaps in k-space, the evolution of the CDW wave vector with R and the shape of the residual metallic pockets. Finally, we give an estimation of the CDW interaction parameters and find that the change in the electronic density of states $n(E_f)$, due to lattice expansion when different R ions are inserted, has the correct order of magnitude to explain the evolution of the CDW properties.

## I. INTRODUCTION

Charge density waves (CDWs) are typical instabilities of the Fermi Surface (FS) in the presence of electron-phonon coupling.[1] They occur when many electrons can be excited with the same q vector of one particular phonon mode at a moderate energy cost, i.e. by keeping these electrons near the Fermi level. The ideal case is when *all* electrons can be excited this way, which implies that all parts of FS can be connected by q to some other parts. This property of the FS is called perfect nesting. For an ideal one dimensional (1D) system, the FS consists of two points, at $-k_f$ and $+k_f$, so that it exhibits by definition perfect nesting at $q=2k_f$. Hence, all 1D systems are subject to CDW transitions, often called Peierls transitions in this case. In real systems, the good nesting properties are usually reduced to some particular regions of the FS, but are nevertheless often sufficient to trigger CDWs (or Spin Density Waves (SDWs), their spin analog) in many quasi-1D, 2D or even 3D systems (a famous example is the SDW of $Cr^2$). In this case, the CDW/SDW gap is expected to open only on the best nested FS parts and the system may remain metallic in its broken symmetry ground state.

Although these ideas were introduced in the 1950s and CDW were heavily studied experimentally since the mid-1970s,[3] they remain a subject of interest for today's research, because of the continued interest in low dimensional systems. Such systems indeed offer fascinating opportunities to study new states of matter, where electronic correlations probably play a major role, because the spatial confinement enhances the probability of interactions. In these systems, CDWs or SDWs are often instabilities competing with more exotic ground states, and they are therefore important to fully characterize. For example, Yao *et al.* used $RTe_3$ to study the competition between checkerboard and striped charge orders, which may be of relevance for comparison with cuprates or nickelates.[4]

Also, the technical progress of Angle Resolved Photoemission (ARPES) in the past 2 decades, has made it possible, in principle, to illustrate very elegantly the impact of the CDW formation on the electronic structure, in a much more direct way than with any other experimental methods. As ARPES produces images of the FS, one can directly examine its nesting properties and compare them with the strength of the CDW gap measured on the different FS parts. Despite this, there are not many examples of CDW systems, where ARPES could be used to illustrate all these points. This is mainly because finding a truly low dimensional system that remains relatively simple is a rather difficult task. As we have seen, quasi-1D systems are the most likely hosts for CDW. However, they also exhibit by nature serious deviations from the Fermi-liquid theory that complicates ARPES analysis. In particular, Fermi edges are usually not well defined, making the definition of gaps more difficult, as in organic conductors,[5] $(TaSe_4)_2I$[6] or $K_{0.3}MoO_3$.[6,7] Also, for a truly 1D system the gap would open homogeneously on the FS, which reduces the interest of a k-resolved probe like ARPES. In the quasi-1D system $NbSe_3$, imperfect nesting gives rise to coexistence between gapped and metallic regions, which could be observed in ARPES[8], despite a rather complicated band structure. Hence, quasi-2D systems appear as a more simple choice for photoemission studies. Transition metal chalcogenides ($1T$-$MX_2$ or $2H$-$MX_2$ with M=Ti, Nb, Ta and X=S, Se, Te) exhibit a variety of quasi-2D CDW behaviors that have been extensively studied with ARPES.[5,9] Despite this, no simple relation between nesting properties and the reported gaps (ususally 10-20meV) could be firmly established[9,10] and the mechanism of the CDW itself is still debated.[11] Let us emphasize that despite the presence of chalcogenides, these systems have little in common with $RTe_3$. Triangular planes of transition metal ions dominate their electronic properties, while these are square planes of Te in $RTe_3$. To conclude this short overview of ARPES in CDW systems, let us mention the case of the surface CDW in In/Cu(001),[12] which exhibits interesting similarities with $RTe_3$, although it is not a bulk phase transition.

$RTe_3$ are quasi-2D metals, where a much clearer situation is encountered. We will argue that they do allow the illustration of the main CDW concepts fairly well and further raise interesting questions on the limits of different CDW models. After the seminal work initiated by DiMasi et al.[13-15], there has been a growing wealth of information about these materials gathered through detailed structural[16,173], STM[16,18], ARPES[19,20], transport[21] and optical studies[22,23]. It was first believed that these compounds always remain in the CDW state, up to their melting point, making the language of phase transition questionable. We have recently revealed a transition to the normal state at 244K in $TmTe_3$ up to 416K in $SmTe_3$ and presumably even higher temperature for lighter R.[17] This definitely qualifies this family as CDW materials and opens new perspectives for a full characterization of the CDW state, including the fluctuations above the transition. In this paper, we will restrict our study to the characterization of the ground state of the light rare-earths, from $LaTe_3$ to $DyTe_3$. For the heavy rare-earth ($DyTe_3$ and above), two successive phase transitions occur,[17] which we do not consider here. The possibility of tuning the CDW properties (the transition temperature, the size of the gap, the wave vector $q_{cdw}$) with R is a rather unique property of this family, which is very useful to discuss the origin of the CDW. A similar variation of the CDW properties can be induced by applied pressure,[23] making it likely that the changes are due to lattice contraction.

The CDW in $RTe_3$ is characterized by large displacements (about 0.2Å)[16] and large gaps in the electronic structure (up to 0.4eV).[19,20,22] This large gap is an advantage for ARPES studies, because it makes it easy to measure its location and changes in k-space accurately. On the other hand, it raises questions about the nature of the CDW and especially whether the traditional weak coupling treatment of the nesting driven CDW would still hold. Indeed, the gap is presumably several times larger than the phonon frequencies involved. In such a situation, a strong coupling model of the CDW could appear more appropriate, where the structural distortion is really the driving force for the transition. The local tendency of Te atoms to form stable chemical bonds would be its starting point. Indeed, a usual Te-Te bond is 2.8Å, whereas the average distance between Te in the planes hosting the CDW is about 3.1Å. Whangbo and Canadell discussed, in the case of $1T$- or $2H$-$MX_2$, similarities and differences between the approach of FS nesting or chemical bonding.[24] The distinction between weak and strong CDWs was investigated by Nakagawa et al. in their study of In/Cu(001).[12] They concluded the CDW in this system was of "dual nature". In this paper, we will show, with unprecedented details for any CDW system, that the predictions of the FS nesting scenario explain extremely well the openings of gaps observed by ARPES in $RTe_3$. Furthermore, the main variations of the CDW properties with R can be well explained by an additional stabilization of the CDW due to the enhancement of $n(E_f)$ through lattice contraction. This gives substantial ground that the electronic energy is, at least, an essential ingredient for the formation of the CDW in $RTe_3$.

Large single crystals of $RTe_3$ were grown by slow cooling of a binary melt.[21] The crystals easily cleave between two Te planes, providing a good surface quality for ARPES. ARPES measurements were mostly carried out at the beamline (BL) 10.0.1 of the Advanced Light Source (ALS), with a Scienta-2002 analyser, an energy resolution better than 20meV and an angular resolution of 0.3°. Other data were acquired at BL 12 of the ALS (Fig. 7, Fig. 12) and BL 5-4 of the Stanford Synchrotron Radiation Laboratory (Fig. 11). All measurements were performed at low temperatures T≈20K.

## II. TIGHT-BINDING MODEL OF THE ELECTRONIC STRUCTURE

In RTe$_3$, Te planes are stacked with R/Te slabs,[25] as sketched in Fig. 1a. Note that we follow the usual convention where the b axis is *perpendicular* to the Te planes (b ≈ 26 Å). The planar unit-cell (a,c) is defined by the R atoms of the slab (orange square in Fig. 1a and 1b). There is a small orthorhombic distortion of this square, which shrinks from a=4.405Å and c=4.42Å in LaTe$_3$ to a=4.302Å and c=4.304Å in DyTe$_3$.[17] The Te atoms in the planes form a nearly square net, but with a square unit (green square) rotated by 45° with respect to the unit cell and with only half the area. Hence, two different Brillouin Zones (BZ) will be convenient to use throughout this paper : a 2D BZ built on the Te square from the plane and the 3D BZ built on the lattice unit cell (see Fig. 1c). We define a*=2π/a and c*=2π/c as unit wave vectors of the 3D BZ.

The band structure was calculated using the linear muffin-tin orbital (LMTO) method for a fictitious LuTe$_3$ composition and a=c=4.34Å.[26] Lu was chosen to avoid the complications associated with the description of f electrons in the local-density approximation. The calculated band structure is shown in Fig. 2a along c* at a fixed k$_x$=0.3a* and k$_y$=0. Eighteen different bands are found between 2eV and –6eV, corresponding to the Te 5p orbitals of the 6 Te per unit cell (2 in the slab and 4 in the planes). However, only 4 bands cross the Fermi level, corresponding to the Te in-plane p$_x$ and p$_z$ orbitals. They are well isolated from other bands over a 1eV window below E$_f$.

The dispersion of these bands can be very well reproduced by a tight-binding (TB) model of the Te plane. We consider only the two perpendicular chains of p$_x$ and p$_z$ orbitals, represented on Fig. 1b in red and blue, with a coupling t$_{para}$ along the chain and t$_{perp}$ perpendicular to the chains. We assume a square net and totally neglect the coupling between p$_x$ and p$_z$. Using the axes of the 3D BZ, this yields the following dispersions.

$$E_{p_x}(k_x, k_z) = -2t_{para} * \cos[(k_x + k_z)*a/2] - 2t_{perp} * \cos[(k_x - k_z)*a/2] - E_f$$
$$E_{p_z}(k_x, k_z) = -2t_{para} * \cos[(k_x - k_z)*a/2] - 2t_{perp} * \cos[(k_x + k_z)*a/2] - E_f$$

This model is plotted on top of the calculated band structure in Fig. 2a, as red and blue lines for p$_x$ and p$_z$, respectively. As the TB bands are constructed for one Te plane, they have the periodicity of the 2D BZ and they have to be folded back with respect to the 3D BZ boundaries to acquire the 3D lattice symmetry. These additional folded bands are shown as dotted lines in Fig. 2a. The Fermi level E$_f$ =-2t$_{para}$sin(π/8) was fixed so that $p_x$ and $p_z$ each contain 1.25 electrons. We assume here that all R are trivalent[15,21] and donate 2 electrons to each Te in the slab and 0.5 electrons to each Te of the 2 planes. We further assume that the out-of-plane Te orbital $p_y$ is completely filled, leaving 2.5 electrons for $p_x$ and $p_z$. The TB parameters were adjusted to reproduce best the calculated band structure, which is reached for $t_{para}$=-1.9eV and $t_{perp}$=0.35eV. It is worth noting that although $|t_{perp}|<<|t_{para}|$, it is much larger than the temperature ($t_{perp} \approx$ 3000K), so that one would not be in a 1D limit, even for an isolated set of chains. As there are two Te planes per unit-cell, the bands in the calculation are doubled and there is a clear "bilayer splitting" between them. In this calculation, spin-orbit couplings were neglected.

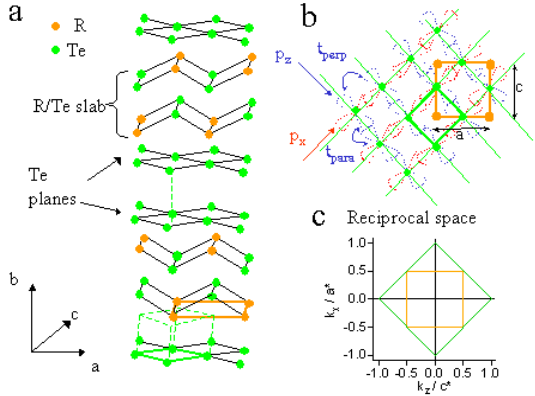

Fig. 1 : (a) Sketch of the RTe$_3$ structure. (b) Sketch of the Te plane (green points) with in-plane p$_x$ and p$_z$ orbitals in red and blue. The 3D (a,c) unit cell is shown as orange square. (c). Sketch of the reciprocal space with the 3D BZ (orange) and the 2D BZ (green) that would correspond to one isolated Te plane.

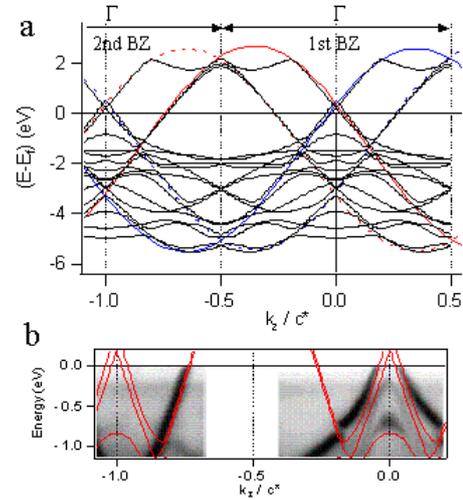

Fig. 2 : (a) Band Structure along c* for k$_x$=0.3a* and k$_z$=0 calculated with LMTO method. Red and blue lines are TB fits for p$_x$ and p$_z$. Dotted lines are folded bands. (b) Zoom-in of the electronic structure near E$_f$ measured with ARPES in CeTe$_3$ at 55eV along the same direction. Red lines are the calculated bands.

Fig. 2b shows the corresponding electronic structure measured with ARPES. The agreement with the calculation for the bands at the Fermi level is very good, except the intensity of the folded bands is so weak that they are hardly distinguishable. As explained in ref. 20, the intensity of folded bands is very generally proportional to the strength of the coupling responsible for the folding. The weak intensity of the folded bands reflects here the small 3D couplings and consequently, the nearly 2D character of these compounds. For the deeper Te bands, we observe some deviations between the calculated and measured bands, more details will be given in part II.C.2.

The excellent description of the electronic structure near $E_f$ with only in-plane Te orbitals suggest a negligible coupling with the R/Te slab. The transport anisotropy is indeed very large, at least a factor 100.[15,21] In this case, one expects that the main consequence of changing R will be a change in bandwidth due to the expansion or contraction of the Te square lattice. Fig. 3 displays the ARPES intensity of the Te bands integrated around the Γ point, for different rare-earths. Their structure is quite similar, confirming the small influence of rare-earth orbitals. The total bandwidth can be estimated by the peak position of the last band. It increases from about 4.25eV for Ce (c=4.385Å) to 4.70eV in Gd (c=4.33Å) and Tb (c=4.314Å). This is in qualitative agreement with the larger overlaps between Te orbitals expected for smaller lattice parameters.

In our calculation, the bandwidth at Γ increases when expanding the lattice from 4.75eV (La, c=4.42Å) to 5.15eV (Dy, c=4.03Å), i.e. by 8%. While these absolute values are a little larger than the experimental ones, the order of magnitude of the expansion is in good agreement. This corresponds in the calculation to a decrease of the density of states at the Fermi level from $n(E_f)$=1.6states/eV/cell (La) to 1.5states/eV/cell (Dy). With the TB model, for $t_{para}$=-1.9eV and $t_{perp}$=0.35eV estimated before for a=c=4.34Å, we calculate $n(E_f)$=1.48states/eV/cell, slightly smaller. This is the same trend as for the bandwidth and suggests that a slightly smaller value of $t_{para}$ might be more appropriate to describe RTe$_3$. We observe that $n(E_f)$ solely depends on $t_{para}$ for realistic values of $t_{perp}$. To reproduce the calculated $n(E_f)$ values, one has to use $t_{para}$=-1.7eV (La) to –1.85eV (Dy). This is a variation of 8%, in good agreement with that of the bandwidth, both calculated and experimentally observed. Therefore, we will use this parameter range in the rest of the paper to model the changes of the electronic structure from La to Dy.

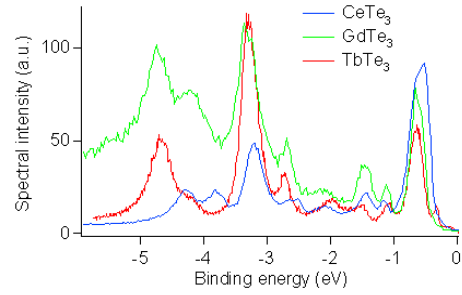

Fig. 3 : Te valence bands integrated around the Γ point, measured for different rare-earths, at a photon energy of 55 eV.

### A. Fermi Surface

The FS expected in the TB model is very simple. It is made out of two perpendicular sets of nearly parallel lines, corresponding to the two chains. They are shown in Fig. 4a in red and blue, for $p_x$ and $p_z$ respectively. With no perpendicular coupling ($t_{perp}$ = 0), the problem would be reduced to that of two perfectly 1D chains, perpendicular to each other, and the FS would consist of two sets of exactly straight lines, exhibiting perfect nesting.[14] The coupling between the chains introduces a deviation from one dimensionality and a curvature of the FS proportional to $|t_{perp}/t_{para}|$. We will show in part III (e.g. Fig. 17) that it is precisely this curvature that makes the nesting imperfect. The orange arrow indicates the best nesting wave vector $q_N$=0.68c*. There are other wave vectors giving better (actually perfect) nesting for $p_x$ or $p_z$, but this one reaches a better compromise by nesting equivalently $p_x$ and $p_z$. The competition between these different wave vectors has been studied by Yao *et al.*[4]

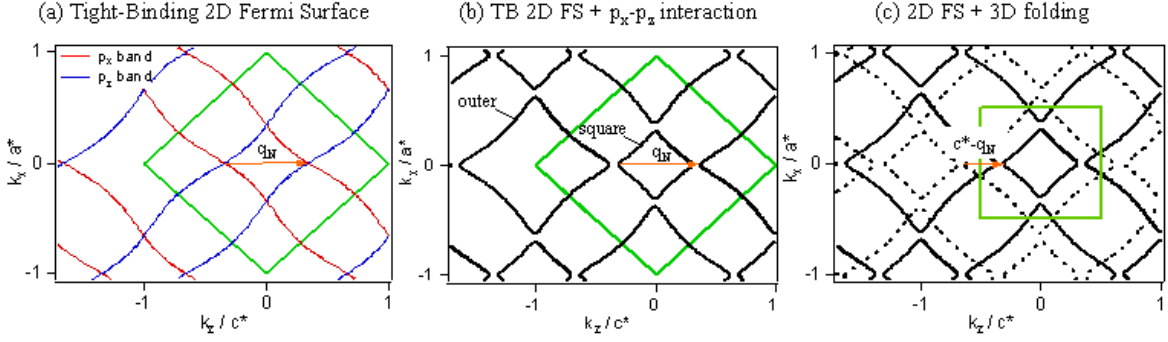

Fig. 4 : (a) Tight Binding FS for ($t_{para}$=-1.7eV, $t_{perp}$=0.35eV) and ($t_{para}$=-1.85eV, $t_{perp}$=0.35eV). The two contours overlap almost perfectly (see text). The green square delimits the BZ corresponding to one Te plane. The orange arrow represents the best nesting wave vector $q_N$. (b) Same as (a) after interaction between $p_x$ and $p_z$ that separates the "square" part of FS from the "outer" part. (c) Same as (b) plus, as dotted lines, the FS contours folded with respect to the 3D BZ limits (green square). The orange arrow represents the equivalent nesting direction, but defined in the 3D BZ : ($c^*-q_N$).

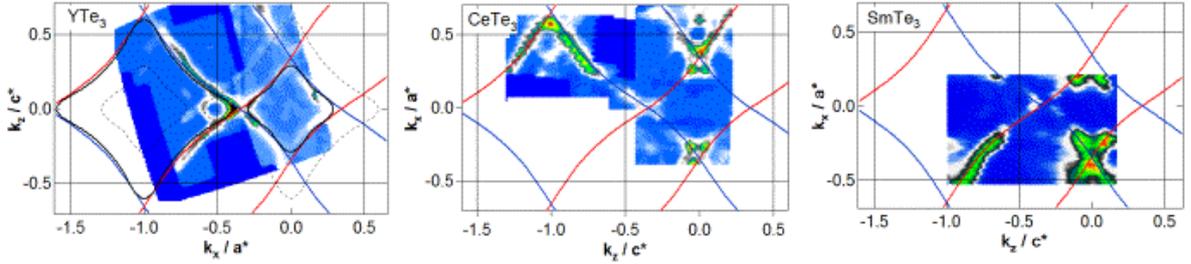

Fig. 5 : Fermi Surface for $YTe_3$, $CeTe_3$ and $SmTe_3$, obtained by integration of spectral weight in a 10meV window around $E_f$. No symmetry operation were applied to the data. Photon energy was 35eV for $YTe_3$, 55eV for $CeTe_3$ and $SmTe_3$, polarization was almost perpendicular to the sample surface. Red and blue lines correspond to the Fermi Surface calculated with the TB model described in the text. Black contours on $YTe_3$ map are guides for the eyes for the shape of square and outer pieces of FS. The suppression of spectral weight around $k_X$=0 is due to the opening of the CDW gap (note that $YTe_3$ map is rotated by 90°).

The TB FS in Fig. 4a is shown for the two sets of parameters corresponding to Dy and La. The two contours are however so close that they cannot be distinguished. This is normal, as the FS area should be proportional to the number of holes in the band, which is kept constant in our model. When the band width changes, the Fermi level readjusts to keep this area constant. Therefore, the FS contour is independent of $t_{para}$, as long as its shape (i.e. $|t_{perp}/t_{para}|$) remains the same. Interestingly, this means that, in this model, the nesting properties remain *exactly* the same throughout the series.

The experimental FS of $RTe_3$ was first measured by Gweon *et al.* in $SmTe_3$.[19] In Fig. 5, we present the FS for different R = Y, Ce and Sm, compared with the predictions of the TB model. FS are obtained by integration of the spectral weight in a 10meV window around $E_f$. The spectral weight is strongly suppressed in a large region around $k_x$=0 (note that the axes are rotated for $YTe_3$). We will show in part III that this is due to the opening of a large CDW gap in these regions. Although the directions of $k_x$ and $k_z$ appear at first quite similar in structure, x-ray measurements have shown that the gap always open along the c axis.[17] Accordingly, we only observe the gap opening along $k_z$. These regions are the best nested ones (see Fig. 17), in those with poorer nesting, the gap does not open at the Fermi level and we observe again intensity at $E_f$. Interestingly, the ungapped regions appear larger in $YTe_3$ and $SmTe_3$ than $CeTe_3$ (this is particularly clear for the ungapped fraction of the square). We will analyse this behavior quantitatively in part III and show that it can be understood from the larger gap of $CeTe_3$.

Clearly, the distribution of the spectral intensity is equally well described by the TB model in the three cases and this was true for all the rare-earths we have measured. As discussed before, this does not give information on $t_{para}$, but rather proves that there are no significant changes in the band filling. This is not a trivial result as it is for example not the case in $RTe_2$.[27] On the other hand, the well-defined curvature of the FS allows us to estimate $t_{perp}$=0.35±0.08eV. This is a totally independent estimation from the previous section, but turns out to be in very good agreement.

The major deviation between the experimental FS

and the TB fit takes place at the crossing between $p_x$ and $p_z$. This is because there is no coupling between these bands considered in our TB model. Fig. 5 shows that, in reality, they do interact and this rounds the FS contours near the crossings. This effect is simulated in Fig. 4b : it creates two different sheets of FS, a small hole-like piece around Γ, called hereafter "square", and a larger electron-like piece mainly in the second BZs, called hereafter "outer". Note that the nesting quality and the wave vector do not change when this interaction is added, because the effect is symmetric on the square and outer FS that are nested into one another.

In the 3D BZ, the folding of the FS gives rise to the dotted black contours of Fig. 4c. Experimentally, Fig. 5 shows that their intensity is always very weak, except near the zone boundaries.[20] The nesting properties are the same, except they would be described by $(c^*-q_N)$ in the 3D BZ.

### B. Dispersion

In Fig. 2, one can check that not only the position of the FS crossings but also the slopes of the dispersions are in very good agreement with the TB model. This value directly depends on $t_{para}$, so that its evolution as a function of rare-earth may give additional insights into the evolution of the electronic structure.

Metallic properties are best measured on the outer part of FS, where the lineshape is more simple (see parts C and D). Near $E_f$, the dispersion is nearly linear and we extract its slope by a linear fit over a 0.2eV window. We observe that the changes in the slope of the dispersion are small along the outer FS and also as a function of R. In fact, the slope essentially depends on the direction in which the dispersion is measured, i.e. the angle α of the detector slits of the analyser with respect to $k_z$ axis.

The TB model again offers a useful guide to understand this evolution. As the bands are essentially one dimensional, the slope of the dispersion is nearly constant when measured along the chain direction (i.e. α=±45°). We define this value as the reference Fermi velocity, which is $V_f = \sqrt{2}at_{para}\sin(k_f a)$. On the other hand, the slope of the dispersion rapidly falls to zero if measured perpendicularly to the chain. This dependence is illustrated on Fig. 6a for the two values of $k_x$ between which the dispersion can be measured reliably on the outer FS. It is nearly a cosine function of α and only the small $t_{perp}$ gives it some $k_x$ dependence. Many different values measured for different samples and/or branches of the FS are reported as color points. The variation with $k_x$ is indeed within the error bar of the measurement and the general trend is the dependence with α. In Fig. 6b, we plot $V_f$ as a function of the lattice parameter c, after correcting for the α dependence. We obtain an average value for all samples $V_f$=10±1 eV.Å. Within the TB model, this corresponds to $t_{para}$=-1.7±0.15eV, a value that corroborates our previous estimation. The variation expected as a function of lattice parameter in the previous paragraph ($t_{para}$=-1.7 to 1.85eV) is shown in Fig. 6b to be within the error bar of the measurement. Let us note that the dispersion is defined here over a rather large energy scale and this analysis does not exclude possible renormalization effects near the Fermi level.

### C. Deviations from the tight-binding model

#### 1. Ce contribution

The smooth changes of the electronic structure as a function of rare-earth is a good indication that they do not play an active role in the electronic properties. However, it would be interesting to clarify the relationship between the localized 4f moments on the rare-earth and the Te band, even if their coupling is weak. We present here results for CeTe$_3$, for which the magnetic susceptibility indicates localized moments of 2.4$\mu_B$, consistent with trivalent Ce that order antiferromagnetically at T$_N$=2.8K.[21,28] There is a mild upturn of the resistivity below 10K, suggestive of a weak Kondo behavior.[21]

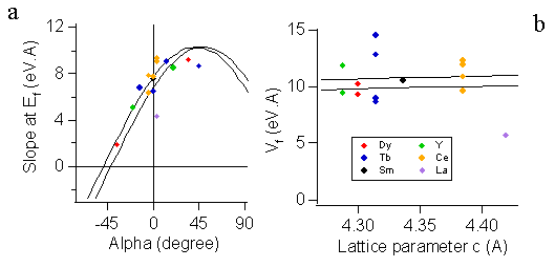

Fig. 6 : (a) Color points : measured values of the slope of dispersion for different RTe$_3$ compounds and different branches of FS, as a function of the angle α of the measurement. Black lines : variation of the slope of the dispersion near $E_f$, calculated in the TB model, for $k_x$=0.6a* and 0.7a*, $t_{para}$=-1.7eV and $t_{perp}$=0.35eV. (b) $V_f$ values corrected for the α dependence as a function of lattice parameter. Black lines are theoretical variations of $V_f$ for $t_{para}$=-1.7eV and –1.85eV.

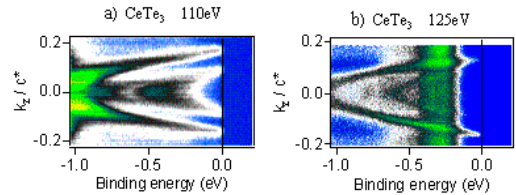

Fig. 7 : Comparison of the electronic structure, measured at T=20K and $k_x$=0.6a*, in CeTe$_3$ at a photon energy of (a) 110eV and (b) 125eV, i.e. respectively off and on the Ce 4d-4f resonance. The two parabola come from the main and folded outer Te bands (see Fig.8), the line at –0.28eV in (b) is attributed to the spin-orbit satellite of the Ce resonance.

In CeTe$_3$, we observe a non-dispersive line at E=-0.28eV throughout the whole BZ. In Fig. 7, we show that it is strongly enhanced at the Ce 4d-4f transition (120eV), indicating it has Ce character. This line is indeed absent in other RTe$_3$ systems. It does not interact strongly with Te bands, as there is no detectable perturbation of the Te dispersion at their crossings.

Generally, one would expect a two peak structure for the Ce spectrum, corresponding to screened and unscreened final states of the hole created through the photoemission process (respectively 4f$^1$ and 4f$^0$).[29] Their positions and relative intensities are very sensitive to the nature of the coupling between localized moments and the metallic band. Such two peaks were observed in CeTe$_2$ at –4eV (4f$^0$) and –1eV (4f$^1$),[30] which is typical of a localized Ce$^{3+}$ in a nearly insulating medium. In CeTe$_3$, the 4f$^1$ peak moves closer to the Fermi level (-0.28eV) and we did not resolve the 4f$^0$ peak from other Te bands at lower binding energies. The 4f$^1$ peak is known to exhibit a spin-orbit splitting of 0.28eV between 4f$^1_{7/2}$ and 4f$^1_{5/2}$ states. This suggests that the satellite line we observe is in fact the 4f$^1_{5/2}$ spin-orbit satellite of a 4f$^1_{7/2}$ peak centred at the Fermi level, but having a negligibly small intensity.[31] This is the situation expected at temperatures higher than the Kondo temperature T$_K$.[29] As this measurement was done at T=20K, this result corroborates the idea that CeTe$_3$ is a weak Kondo system, with T$_K$<<20K.

### 2. Bilayer splitting

One thing neglected in the TB model is the coupling between the Te planes that gives rise to the bilayer splitting (see Fig. 2). Fig. 8 shows the FS of LuTe$_3$ obtained with the LMTO method. The shaded area corresponds to the gapped area. The amplitude of the bilayer splitting is indicated as color scale. It changes quite strongly along the FS, it is larger in the square ($\delta$=0.03c∗) than in the outer part (typically less than $\delta$=0.01c∗). The typical full width of our spectra at half maximum are found between 0.02 and 0.03c∗. Consequently, bilayer splitting is usually not resolved on the outer Fermi Surface (except near the corners) but it is in the square. In Fig. 9 and 10, we give examples of the typical lineshapes observed in these two parts.

In the square, we typically observe two lines reaching for E$_f$. Fig. 9 gives two examples at different heights in the square, one metallic (Fig. 9a) and one gapped (Fig. 9b). They form a sort of inverted V-like shape that is tempting to attribute to the bilayer splitting, although, in the calculations, the two lines are more parallel, at least near E$_f$. The relative intensities of the two lines are very sensitive to the photon energy, they are quite different at 55eV (Fig. 9a) but nearly equal at 35eV (Fig. 9b). Such oscillations in the intensity of bilayer split bands with photon energy were also observed in the well studied case of Bi$_2$Sr$_2$CaCu$_2$O$_{8+\delta}$.[32] The curvature of the outside band seems to be due to the crossing with the folded band at –0.8eV (see also Fig. 2), but it appears more pronounced experimentally (especially in Fig. 9a).

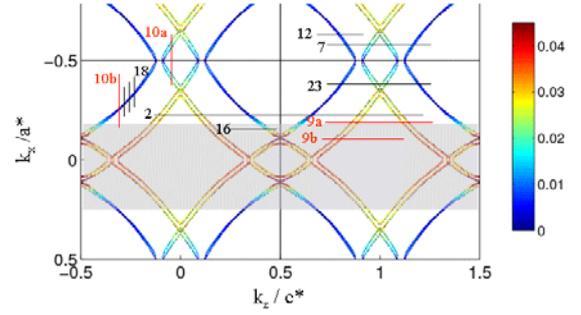

Fig. 8 : LuTe$_3$ FS calculated with the LMTO method. The color scale indicates the amplitude of bilayer splitting. The shaded area corresponds to the part of FS gapped by CDW. The red lines correspond to cuts used to give example of bilayer splitting, in Fig. 9 and 10. Other lines correspond to cuts shown throughout this paper in figures with number as indicated.

Other bands are present within the square corresponding to different Te orbitals. The agreement with the calculation is not as good as for the near E$_f$ bands, which is probably a consequence of the approximations used in the LDA. We frequently observe shifts of deep Te bands to lower binding energies. This is the case in Fig. 9b, where a band with a distinctive two lobe structure seems pushed along the dotted arrow by about 0.5eV. These shifts seem common, as a similar trend was reported in LaTe$_2$.[33]

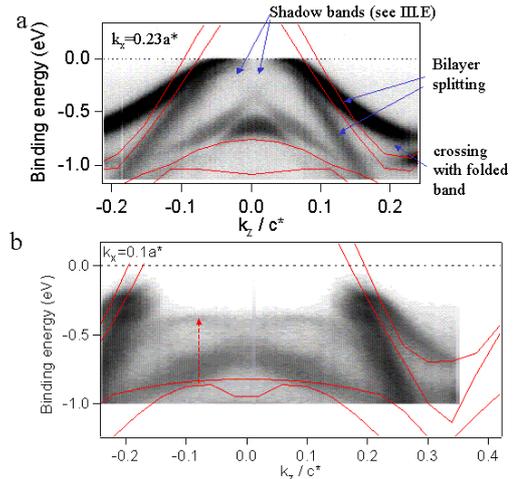

Fig. 9 : Near E$_f$ band structure in the square. (a) Measured in CeTe$_3$ with photon energy of 55eV at k$_x$=0.23a* and (b) in TbTe$_3$ with 35eV and k$_X$=0.1a*. Red lines are calculated dispersions.

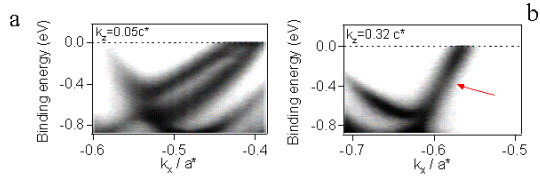

Fig. 10 : Near $E_f$ band structure along the outer FS measured in YTe$_3$ in the 2$^{nd}$ BZ (see Fig. 8) at 35eV, for (a) $k_z=0.05c*$ and (b) $k_z=0.32c*$.

Fig. 10 shows a typical situation for the outer FS. Near the corner ($k_z=0$), the bilayer splitting is maximum. In Fig. 10a, the two lines are well resolved, they have almost the same intensity because the photon energy is 35eV, and the bilayer splitting is 0.025a* near $E_f$. At higher $k_z$, the bilayer splitting rapidly decreases and is not resolved anymore along most of the outer FS, as shown in Fig. 10b for $k_z=0.32c*$ (see position in Fig. 8). The linewidth of the main line is here $\delta\nu=0.025a*$, indeed quite larger than the calculated splitting at this position, $\delta=0.005\pm0.002a*$ (the error bar includes differences depending on the details of the calculation). It is also unlikely that this width is dominated by the bilayer splitting, as the lines are not narrower when it is resolved, as in the square ($\delta\nu=0.024c*$ in Fig.9a) or the outer corner ($\delta\nu=0.035a*$ in Fig. 10a).

### 3. 2D character

An interesting issue is the strength of the 3D couplings in these quasi-2D systems, which we have implicitly neglected so far. Although the transport anisotropy is large, these systems remain metallic along the b-axis[15,21] implying sizable hybridization in the perpendicular direction. 3D couplings often complicate the analysis of the photoemission spectra and might be responsible for a residual broadening of the spectra.[34] Their order of magnitude is therefore important to evaluate.

Fig. 11a shows the dispersion of Fig. 2 calculated for different values of $k_y$. The value of $k_\perp$ in an ARPES measurement is not known precisely, because it is not conserved at the surface crossing. It depends on the photon energy $E_{ph}$ ; at the $\Gamma$ point, $k_\perp = 2m/\hbar^2 \sqrt{E_{ph}+V_0}$, where $V_0$ is an inner potential adjusted experimentally.[35] One can therefore expect a line to *shift* and/or *broaden* as a function of the photon energy on the energy scale of the perpendicular dispersion. The relative amount of broadening and shift will depend on the lifetime of the photoelectron in the final state.[34] Indeed, the linewidth can be written in simple cases as $\delta\nu=\Gamma_i+v_{i\perp}/v_{f\perp}\Gamma_f$, where $\Gamma_i$ and $\Gamma_f$ are the lifetimes in the initial and final states, $v_{i\perp}$ and $v_{f\perp}$ are the slopes of the perpendicular dispersion in each state.[36] One is typically interested in $\Gamma_i$, but it is usually masked by the larger $\Gamma_f$ (the final state having a much higher energy, of the order of the photon energy), unless $v_{i\perp}$ is very small.

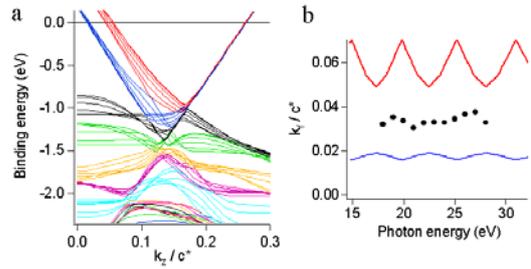

Fig. 11 : (a) Calculated band dispersion at $k_x=0.3a*$, for different $k_y$ values. (b) Black points : $k_f$ position measured in CeTe$_3$ for the square as a function of photon energy. Solid lines : perpendicular dispersion expected theoretically on the square for the red and blue bands of pannel (a).

For the outer bands, the dispersion is fairly independent of $k_y$, as expected for a good 2D metal. The total spread of $k_f$ as a function of $k_y$ is $\delta_{3D}=9.10^{-4}c*$ and $\delta_{3D}=2.10^{-4}c*$, for the two split bands, obviously totally negligible compared to the width of $\delta\nu=0.02$-$0.03c*$. The perpendicular dispersion is larger in the square, as could already be anticipated from the larger bilayer splitting, indicative of stronger transverse couplings. Fig. 11b displays the variations of $k_f$ expected theoretically, for the red and blue bands forming the square (solid lines). They are plotted as a function of photon energy, assuming a typical value $V_0=10$eV [34,35] in the previous formula of $k_\perp$. They are compared with $k_f$ values measured in CeTe$_3$ at different photon energies (black points). Note that the bilayer splitting is not resolved in the square near $E_f$ (Fig. 9a). We do not observe oscillations in the position measured for $k_f$ and the variation is less 0.007c*. Although the linewidth is of the same order of magnitude as the calculated perpendicular dispersion for the red band, it is unlikely that it is dominated by $\Gamma_f$, since the linewidth is very similar for the outer band, where $v_{i\perp}$ is reduced by a factor 20. We conclude that there are no obvious contribution of 3D couplings detectable in our spectra.

**D. Spectral lineshape**

Because of its simple, well understood and nearly 2D electronic structure, RTe$_3$ offers a favorable situation to extract detailed information about the electronic self-energy from the ARPES lineshape.[32,35] Since examples of "simple" low dimensional systems are rare, this deserves attention. Fig. 12 exemplifies typical lineshapes along the outer FS, where the band is well separated from other bands, the bilayer splitting is minimum and 3D couplings are negligible. They have low backgrounds both for cuts taken at constant energy (MDC) and momentum (EDC), which is another advantage for line fitting. Yet, the linewidth (Fig. 12c) appears rather large and does not exhibit a significant decrease near E$_f$, which would be the fingerprint of a Fermi liquid.[37,38] We note however that the Fermi step is very well defined (see also Fig. 21), so that this case is completely different from that of a "bad metal", where broad lines are associated with a low weight at the Fermi level, due to strong correlations, low dimensionality and/or polaronic effects.[5] It is worth emphasizing this point, because the opening of the CDW gap along c* gives an effective 1D character to the problem, which raises the question of possible 1D features in the physics of RTe$_3$. A power law behavior of the optical conductivity was recently attributed to the formation of a Tomonaga-Luttinger liquid.[22] We do not observe equivalent effects in the ARPES lineshapes.

The behavior of the width in Fig. 12 is typical of that found in all RTe$_3$ systems we have investigated. Most of the changes we observed as a function of binding energy could ultimately be attributed to crossings with other lines (such as the Ce satellite at –0.28eV) or weak CDW shadow bands (red arrow in Fig. 10b, see III.B). The MDC linewidths are comprised between $\delta v_a$=0.03 and 0.05 Å$^{-1}$ for different samples and/or different cleaves. This translates to a rather large energy width for EDC spectra ($\delta v_e$= $\delta v_a$*V$_s$ = 0.2-0.3eV, where V$_s$ is the slope of the dispersion), but this is essentially due to the fact that V$_s$ is about one order of magnitude larger here than in many reference systems, such as cuprates[32] or 1T-TiTe$_2$.[38]

If taken at face value, the MDC width would correspond to a mean free path $l$=2/$\delta v_a$=40-60 Å, which seems rather small for these good metals where quantum oscillations have been observed.[39] As transport and ARPES lifetimes are different, this comparison is only qualitative. We have seen that the bilayer splitting and 3D couplings should be rather negligible contributions on this scale. On the other hand, the angular resolution was usually set to $\delta k_{res}$=0.3°, which is smaller but not negligible compared to these linewidths (it corresponds to 0.02Å$^{-1}$ at 50eV). For these systems with large Fermi velocities, the angular resolution is indeed a much stronger constraint than energy resolution ($\delta E_{res}$ was typically 10-20meV, much smaller than V$_s$ * $\delta k_{res}$). We did not observe a large improvement when using a higher resolution mode of the analyser (0.1°), which rules out that it is entirely a resolution problem. We believe that other types of extrinsic angular broadening could be a problem, for example a distribution of angles at the sample surface. Although these samples cleave very well, giving smooth and shiny surface, they often exhibit curved surfaces, which could limit our effective resolution. ARPES experiments with ultra small spot may be able to clarify this issue. Alternatively, impurities at the surface could reduce the mean free path we measure.

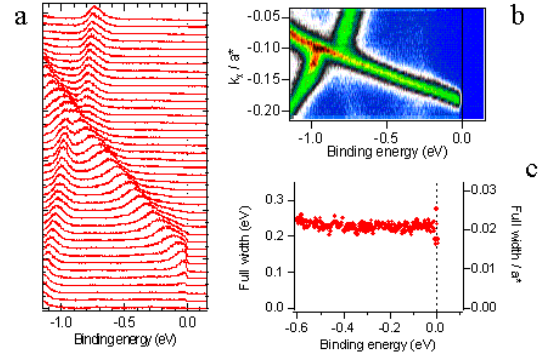

fig. 12 : Lineshape analysis in CeTe$_3$ at 55eV and 20K for a cut of the outer FS at k$_x$=0.62c*. (a) EDC stacks. (b) Image plot. (c) Full width at half maximum of MDC spectra (the value in energy is obtained by multiplying with the slope of dispersion).

**III. CDW PROPERTIES**

Fig. 13 and 14 summarize the main results of our ARPES investigation of the RTe$_3$ CDW properties. The evolution of the gap in k-space is shown in Fig. 13 for different compounds. We find identical gaps on the square and outer FS pieces for a given k$_x$ value, therefore the gap is plotted as a function of k$_x$. The gap is maximum at k$_x$=0 and decreases to zero for a value k$_x^0$ comprised between 0.18 and 0.28 a*. This qualitative behavior is the same for all rare-earth we have studied (Dy, Tb, Gd, Sm, Ce, La) and Y. On the other hand, there are significant quantitative changes in the maximum gap value and in k$_x^0$ as a function of R. Fig. 14 displays the maximum gap value as a function of the lattice parameter. It is defined with respect to the leading edge of the spectra (see part III.D). In $SmTe_3$, our value (280meV) corresponds well to the 260meV first measured by Gweon *et al.*.[19] The increase is roughly linear as a function of lattice constant, with maybe an exception for $YTe_3$, which is the only non rare-earth compound.

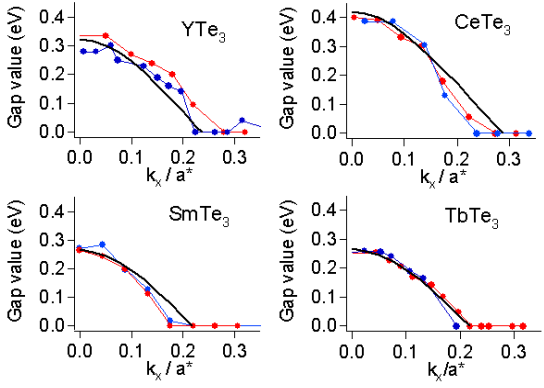

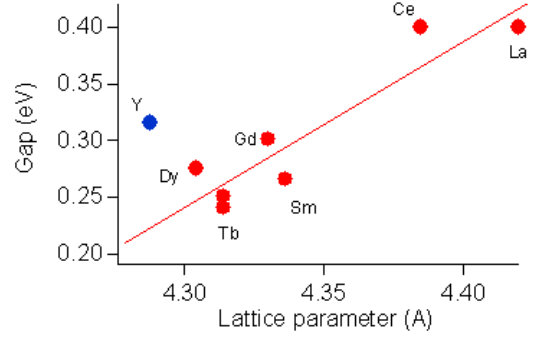

Fig. 13 : k-dependence of the gap along the Fermi Surface for the square part of FS (red points) and outer part (blue points). $k_X$ is used as implicit parameter for the position on FS. The black line describes the decrease of the gap expected because of the imperfect nesting away from $k_x=0$ (see part III.D).

Fig. 14 : Maximum gap value for different rare-earths and $YTe_3$ (blue point), plotted as a function of the lattice parameter c. The straight line is a guide for the eyes.

The change of the gap along the FS is an essential feature to understand the origin of the CDW. We will show that the model of a nesting driven sinusoidal charge density wave allows to explain in details the location of the gap in k-space. This model also implies the presence of metallic pockets and shadow bands in the regions that are not perfectly nested, which we observe and will discuss. The increase of the gap with lattice parameter is another way to investigate the nature of the CDW. The basic idea is that the larger $n(E_f)$ caused by the lattice expansion should stabilize the CDW.[13] We give a quantitative discussion of this phenomenon, which leads to estimation of the important CDW parameters, such as the electron-phonon coupling and the relevant phonon frequencies.

### A. Interacting band structure in the CDW state

We introduce here a simple theoretical model to describe the modification of the band structure in the CDW state. The electron-phonon coupling responsible for the CDW is described by the following hamiltonian.

$$H = \sum_k \varepsilon_k c_k^+ c_k + \sum_{k,q} g_q\, c_{k+q}^+ c_k (a_q + a_{-q}^+)$$

where $g_q$ is the electron-phonon coupling strength for the wave vector q, $c_k^+$ ($c_k$) are creation (destruction) operators for electrons and $a_q^+$ ($a_q$) for phonons.

In the CDW state, a static distortion takes place for a wave vector $\pm q_{cdw}$, implying $\langle a_{\pm q}\rangle = \langle a^+_{\pm q}\rangle \neq 0$ at this wave vector. This creates a coupling $V = 2g_{q_{cdw}} \langle a_{q_{cdw}}\rangle$ between states $|k\rangle$ and $|k \pm q_{cdw}\rangle$. New wave functions allowing for the admixture of these states have to be defined with the form :

$$|\psi_k\rangle = u_{k-q_{cdw}}|k-q_{cdw}\rangle + u_k|k\rangle + u_{k+q_{cdw}}|k+q_{cdw}\rangle$$

where the coefficients are solutions of the matrix

$$\begin{bmatrix} \varepsilon_{k-q_{cdw}} & V & 0 \\ V & \varepsilon_k & V \\ 0 & V & \varepsilon_{k+q_{cdw}} \end{bmatrix}.$$

Here, we truncate the interaction at the first harmonic, although, in principle, all harmonics $n.q_{cdw}$ should be included.

The dispersion of these new wave functions are shown in Fig. 15a, with a size of the markers chosen as $|u_k|^2$. Wave functions are essentially unchanged away from the crossings between $|k\rangle$ (solid line) and $|k \pm q_{cdw}\rangle$ (dotted lines), i.e. $u_k$ =1 or 0. At the crossing, a gap of amplitude 2V opens and some weight is distributed on the translated parts of the band structure (dotted lines), which are called *shadow bands*. Their intensity is proportional to V and decreases very fast away from the crossing. This is better seen in Fig.15b, where the part near the Fermi level is emphasized.

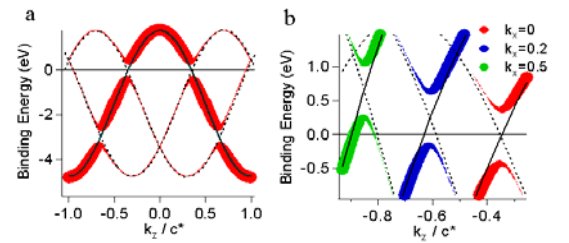

Fig. 15 : (a) Red points : sketch of the dispersion in the CDW state at $k_x=0$, calculated for V=0.4eV and $q_{cdw}=0.32c^*$. The size of the points is proportional to the spectral weight $|u_k|^2$ (see text). The solid black line is the original dispersion from the TB model, dotted lines are translated by $c^*\pm q_{cdw}$. (b) Zoom near the Fermi level for the same calculation and different $k_x$ values.

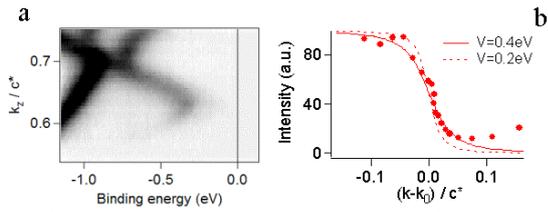

Fig.16 : Shadow band in the gapped state. (a) Band from the outer FS, measured along $k_z$ at $k_x=0.1a^*$ in $CeTe_3$. (b) Red points : intensity of this band as a function of k (zero is the position of the crossing). It is extracted by a fit of the EDC spectra with a gaussian and a parabolic background. Theoretical variation for this weight is also shown for V=0.4eV (as solid line) and V=0.2eV (as dotted line).

An example of the typical resulting shape of the dispersion in the gapped state is shown in Fig.16. The band "turns away" from the Fermi level after reaching a maximum, corresponding to the gap value (here 330meV) and its intensity rapidly vanishes. The observation of such a shape is in fact the best proof that the band is indeed gapped versus its intensity would be accidentally reduced near $E_f$ by matrix element effects. In Fig. 16b, we display its intensity as a function of the distance with respect to the crossing between the main and translated bands. The qualitative variation of this intensity is in good agreement with the expectations of the theoretical model. Two theoretical variations are given for V=0.4eV (solid line) and V=0.2eV (dotted line). However, a quantitative comparison remains difficult, because, in many cases, the photoemission intensity is modulated by matrix element effects that partially mask the intrinsic variation.

Fig. 15b summarizes the evolution expected as a function of the degree of nesting of the FS. When the FS is perfectly nested, the crossing takes place by definition at the Fermi level. The gap measured by ARPES with respect to $E_f$ is maximum and equal to V. This is the case for $k_x=0$ represented in red on Fig. 15b. When the perfect nesting is lost ($k_x>0$), the crossing takes place above $E_f$ and the apparent gap at $E_f$ decreases (see $k_x=0.2$). Eventually, when the crossing takes place at an energy higher than V above $E_f$, the band crosses again the Fermi level and remains metallic (see $k_x=0.5$). In this case, two crossings should in fact be observed at $E_f$, with large and small weights, forming the two sides of a metallic pocket. These theoretical pockets are shown in Fig. 17 for the simplified case of an isolated $p_x$ band. The side with large weight follows the original FS and the other side the FS translated by $c^*\pm q_{cdw}$. Note that we use here the usual definition of $q_{cdw}$ in the 3D BZ, which in fact nests the main and folded FS (see Fig. 4). We however show translated bands by $c^*\pm q_{cdw}$, which is equivalent but connects bands from the same Te plane. The extension of the metallic pocket sensitively depends on the strength of the gap, it is shown for V=0.2eV in Fig. 17a and V=0.4eV in Fig. 17b.

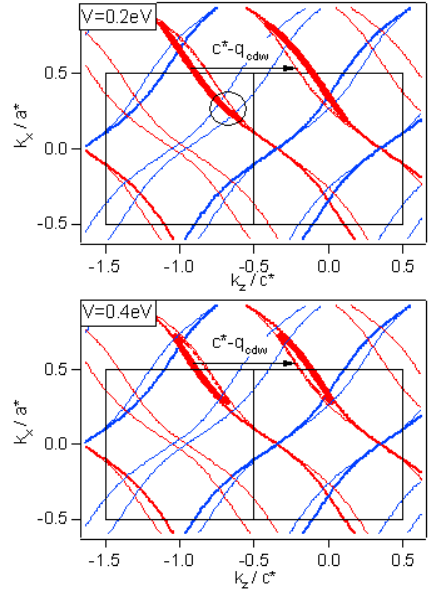

Fig. 17 : Metallic pockets from the $p_x$ band expected after CDW reconstruction, for two different values of V. The interaction with $p_z$ is neglected. Thick red and blue lines correspond to original FS for $p_x$ and $p_z$. Thin lines are FS translated by $c^*+q_{cdw}$ or $c^*-q_{cdw}$. The size of the markers is proportional to the spectral weight.

### B. Value of the interaction parameter V

In the case represented in Fig. 17, the CDW vector has been chosen, so that the original FS and its translated exactly overlap on the corner of the square. In the corresponding band structure (Fig. 15), the Fermi level at $k_x=0$ lies in the middle of the full interacting gap 2V. However, as ARPES only measures occupied states, one cannot directly observe this full interacting gap and this leaves an ambiguity on the size of V. When $k_x$ increases, the symmetry is such that the crossing between the main and translated bands always takes place *above* the Fermi level (Fig. 15b), so that the full interacting gap is still eluding measurement.

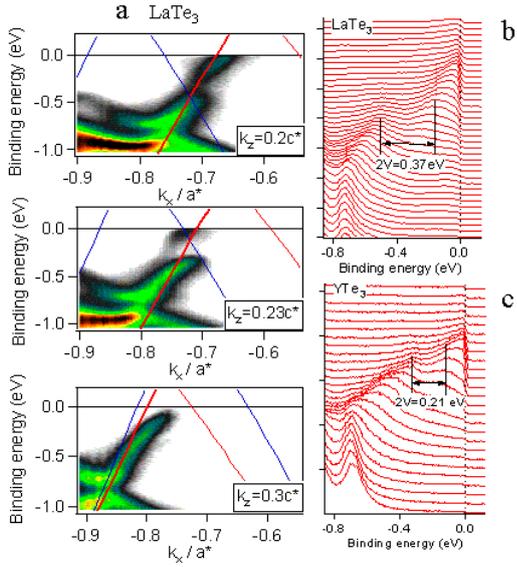
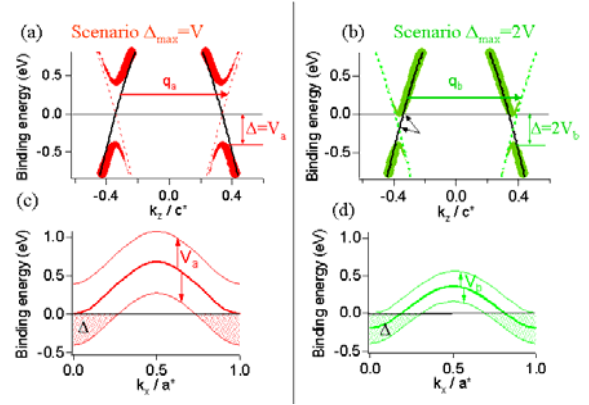

Fig. 18: (a) Dispersion along the outer FS in LaTe$_3$ for three $k_z$ values (h$\nu$=35eV). The TB p$_x$ band is shown as thick red line and CDW shadow bands as thin lines. (b) Detail of the dispersion in LaTe$_3$ at $k_z$=0.2c*. (c) Equivalent situation for YTe$_3$ at $k_z$=0.32c* (Fig. 10b).

Fig. 19 : (a) and (b) : Black line is the TB band at $k_x$=0, dotted lines are translated by ±q, with $q_a$=0.68c* and $q_b$=0.733c*. V is chosen to get the same gap in the two cases : $V_a$=0.4eV and $V_b$=0.2eV. (c) and (d) : Thick line is the position of the crossing between the original and translated bands as a function of $k_x$, for the values of q and V corresponding to (a) and (b). Thin lines are translated by ±V. Dotted areas represent the gapped regions.

A closer inspection of Fig. 17 reveals another type of crossing between main and translated bands, indicated by a circle. These bands belong to different orbitals, they do not nest the original FS, but they link $|k\rangle$ and $|k+q_{cdw}\rangle$ electrons, and are therefore also subject to the interaction V. Their crossing turns out to take place below E$_f$, which allows to directly measure 2V. Indeed, we see these bands interact with each other in our data, as shown in Fig. 18. In Fig. 18a, a strong "break" is observed in the dispersion of LaTe$_3$, at a position corresponding well to that expected for such a crossing. The shadow band intensity is however so weak that one could not guess there is a crossing there without the TB simulation. The lineshape is detailed in Fig. 18b, showing the opening of a break with peak-to-peak distance 2V=0.37eV. The same behavior is observed for YTe$_3$ in Fig. 18c (same data as Fig. 10b). Interestingly, the break is smaller in this case (2V=0.21eV). This is consistent with the smaller gap in YTe$_3$ ($\Delta_{max}$=0.33eV) than in LaTe$_3$ ($\Delta_{max}$=0.4eV). On the other hand, we observe here $\Delta_{max}\approx$2V rather than $\Delta_{max}$=V expected in Fig. 15.

One would obtain $\Delta_{max}$=2V, if the Fermi level at $k_x$=0 was set at the top of the gap rather than in its middle. Fig. 19 compares the band structure at $k_x$=0 for the two scenarios, $\Delta_{max}$=V (Fig. 19a) and $\Delta_{max}$=2V (Fig. 19b). In the second case, $q_{cdw}$ is chosen a little longer to allow the main band and its translated to cross at –V. As shown in Fig. 19c and 19d, the gap observed at the Fermi level by ARPES as a function of $k_x$ would be very similar, although V would be different by a factor two.

One important difference between these two scenarios concerns the *conservation of the number of occupied states*. For the common case of a homogeneously gapped FS, E$_f$ has to be in the middle of the gap to conserve the number of electrons. However, the situation is quite different here, because of the imperfect nesting. As soon as the crossing between the original band and its translated band takes place away from the Fermi level, there is a different number of occupied states in the gapped and metallic state. In the case of Fig. 19b, for example, the states above the crossing position (this region is delimited by black arrows) were occupied in the metallic state, but are empty in the CDW state (the weight in the shadow band is just shifted from the main band, so that this does not change the counting of occupied states). This roughly corresponds to a loss of V/V$_f$*n(k) states, n(k) being the density of k states and V$_f$ the Fermi velocity. More generally, the imbalance is just proportional to the position of the crossing in energy. It is easy to see in Fig. 19d that the loss of states near $k_x$=0 will be compensated by a gain of states for higher $k_x$. On the contrary, in Fig. 19c, there is no such compensation and there would be a significant excess of electrons in the CDW state, which is obviously not self-consistent and requires a shift of E$_f$. Direct calculations of the density of states in metallic and gapped states (see part III.F) shows that the Fermi level moves very close to the position of scenario (b) to conserve the number of electrons. This also maximizes the gain of electronic energy for a given value of V. This gives a natural explanation for the choice of the situation $\Delta_{max} \approx 2V$ observed in Fig. 18.

In principle, probes that also measure unoccupied states, like STM, could directly confirm or infirm this scenario. However, they are not k-resolved and, as there is a large distribution of gaps, the analysis is not straightforward.[18]

### C. Value of $q_{cdw}$

Another consequence of the scenario (b) is that $q_{cdw}$ should change with V. This is an interesting point as a change of $q_{cdw}$ is indeed observed experimentally. In scenario (a), $q_{cdw}$ is fixed by the size of the FS to $2k_f=0.68c*$ at $k_x=0$ (in the 3D BZ, $q_{cdw}=1-2k_f=0.32c*$). As seen in part II.A., this size should not and does not change with R. On the contrary, in scenario (b) $q_{cdw}$ is chosen for the crossing to take place at –V for $k_x=0$ and therefore systematically changes with V.

Satellite positions were observed by TEM[13] to increase from $q_{cdw}=0.27c*$ (for $c \approx 4.4$Å) to $0.31c*$ (for $c \approx 4.3$Å). Our recent x-ray measurements measured precisely satellites at $q_{cdw}=0.296c*$ in TbTe$_3$ ($a=4.312$Å).[17] We report in Fig. 20 values we have measured for different rare-earth.[17] To obtain a crossing at E=–V for $k_x=0$, one gets from the TB model :

$$1 - q_{cdw} = \frac{2}{\pi} Arc\cos\left(\frac{V - E_f}{2(t_{para} + t_{perp})}\right)$$

Fig. 20 illustrates the changes of $q_{cdw}$ expected in this scenario for the two extreme values of $t_{para}$ determined in part II. The change with $t_{para}$ is negligible in front of that with V. The agreement with the experimental data is quite spectacular, although the absolute values are slightly larger. This supports this model as the basic origin of the variation of $q_{cdw}$. Indeed, there is no variation of $q_{cdw}$ with R expected from our study of the nesting properties of the FS in RTe$_3$. If small variations exist as a function of R, they have to be restricted to the error bar of $t_{perp}$ or to subtle changes in bilayer splitting, perpendicular couplings, or interaction between $p_x$ and $p_z$ (arising for example from slightly different orthorhombicities). Such effects are probably important for a complete description of the CDW. The existence of two successive transitions in heavy rare-earth compounds[17] or the deviation of YTe$_3$ from the general trend of RTe$_3$ (Fig. 14) are proofs that there are subtleties not captured by the TB model. However, they are probably not able to overcome the strong trend exposed in Fig. 20.

### D. k-dependence of the gap

We have seen that a variation of the gap along the FS is a natural consequence of the nesting driven CDW, when the FS exhibits imperfect nesting, even when the interaction V itself is isotropic. Conversely, the distribution of the gap over the FS directly informs about the direction of the nesting. For example, the CDW wave vector has to be parallel to c* to explain the opening of the same gap on the square and outer FS at one $k_x$ value (Fig. 13). Such a direction is fully consistent with x-ray and TEM studies.[13,16,17] We have also already seen that the experimental values of $q_{cdw}$ closely correspond to the best FS nesting wave vector.

We now study the k-dependence of the gap quantitatively. Fig. 21a gives examples of leading edge spectra in SmTe$_3$ at different $k_x$ values used to determine the gap in Fig.13. The Fermi edge (black line on Fig. 21a with a width of 30meV, corresponding to the experimental resolution) is very clear, making the determination of the metallic zones unambiguous. When the gap opens, the Fermi edge is preserved, but a double step is commonly found, especially on the outer part. We believe that, rather than a distribution of gap values, this is due to the bilayer splitting. Fig. 21b shows data acquired at 35eV instead of 55eV in Fig. 21a (although in LaTe$_3$ instead of SmTe$_3$). The structure near $E_f$ resembles a double peak, with two different gaps, rather than a line with shoulder. This could be explained by the variation of intensity of the bilayer split bands with photon intensity described in II.C.2. In all cases, we have aligned the Fermi edge on the more deeply gapped part. We note that slightly different gap values were reported for the two parts at a same $k_x$ value in ref. 19. This could be due to a different treatment of this bilayer splitting or to a lower accuracy of this first measurement done only in 1st BZ, where the intensity of the outer part is very weak. In all investigated cases, we have observed identical gaps on the two parts, within experimental accuracy.

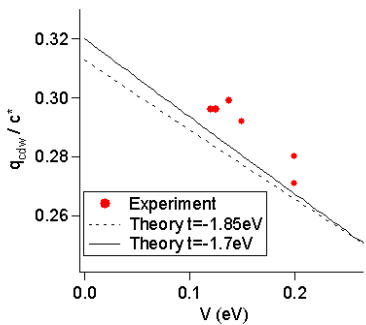

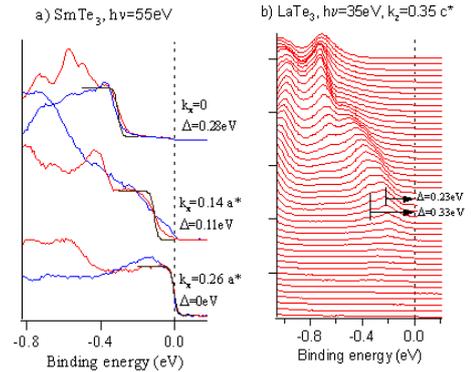

Fig. 20 : Red points : experimental values of $q_{cdw}$ measured by x-ray (ref 17 and 16 for CeTe$_3$), as a function of V=$\Delta$/2 ($\Delta$ is taken from Fig. 14). Black lines : Theoretical variation for $q_{cdw}$ as a function of V (see text), for $t_{para}$=-1.7eV and $t_{para}$=-1.85eV.

Fig. 21 : (a) Leading edge spectra in SmTe$_3$ at different $k_x$ values and hν=55eV on square part (red) and outer part (blue). The black line is a fitted Fermi step. (b) EDC spectra on the outer part in LaTe$_3$ at 35eV and $k_z$=0.35c*.

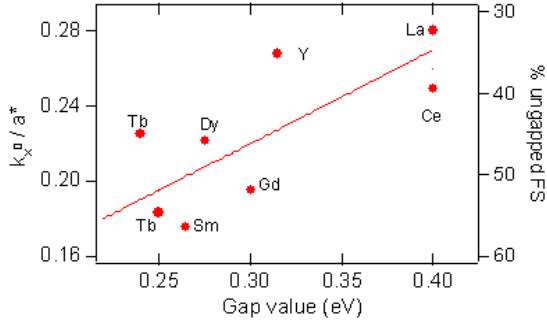

Fig. 22 : $k_x$ value at which the gap on FS becomes zero as a function of the maximum gap value. Samples are the same than in Fig. 14. Straight line is the position of $k_x^0$ calculated as a function of the gap value in the TB model. Right axis indicates the corresponding fraction of the FS contour that remains metallic.

Knowing the value of the CDW wave vector, it is straightforward to deduce the k-dependence of the gap expected within the TB model from the variation of nesting. The principle is that of Fig. 19d and can be calculated for any value of V. We report on Fig. 13 these variations as a black line, with values for $q_{cdw}$ taken from the formula of III.C. The agreement is very satisfying, although the measured gap seems to fall to zero a little faster in the measurement than in the theory. A natural consequence of this simple model is that the position $k_x^0$ at which the gap becomes zero, depends on the maximum gap value. We calculate a linear variation of $k_x^0$ for a gap between 0.2 and 0.4eV, which is shown in

Fig. 22 and fits well with the experimental values, despite some scattering in data points. We conclude that the general behavior of the gap opening directly results from the nesting properties of the FS.

### E. Metallic parts of the electronic structure

A unique feature of RTe$_3$ is the possibility of directly observing by ARPES the reconstructed FS, as we have discussed in ref. 20 and will develop here. Whereas shadow bands are commonly seen in the gapped state[8,12] (like in Fig.16), they are rarely detectable in metallic situations (like $k_x$=0.5 in Fig. 15b), because their intensity when they re-appear at $E_f$ is already very small. However, observing these lines gives detailed information about the deviation from nesting and allow to fully characterize the CDW state.

In Fig. 9a, shadow bands were visible within the metallic part of the square, although they are rather weak. Fig. 23 gives two examples of shadow bands near the junction between square and outer parts (see position in Fig. 8), for CeTe$_3$ and TbTe$_3$, i.e. for large and small V. Six bands cross the Fermi level in this area, $p_x$ and $p_z$, their 2 shadow bands and their 2 folded bands. The dotted red and blue lines in Fig. 23a correspond to the dispersion of $p_x$ and $p_z$ in the CDW state (i.e. including the shadow band). Note that for TbTe$_3$, the asymmetry in $k_z$ is due to a 7° misalignment of the sample. Although weak, the shadow bands are clearly detected. MDC cuts near $E_f$ (Fig.23c and 23d) allow us to estimate their intensity to 12% of the main line in CeTe$_3$ and 3% in TbTe$_3$. Theoretically, one expects in this region 8% at V=0.2eV (CeTe$_3$) and 3% for V=0.14eV (TbTe$_3$), in reasonable agreement with our experimental estimation.

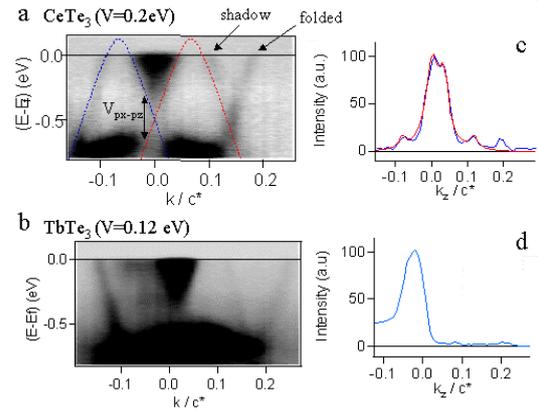

Fig. 23 : (a) Dispersion in CeTe$_3$ along c* at $k_x$=0.38a* and hv=55eV. Dotted red and blue lines are dispersion for $p_x$ and $p_z$ in the CDW state. (b) Dispersion in TbTe$_3$ along a direction at 7° with respect to c*, at $k_x$=0.35a* and hv=55eV. (c) and (d) MDC cut near $E_f$ corresponding to (a) and (b). In (c), the red line is a 3 lorentzians fit.

Note that these predicted intensities are still quite small, so that it is clear that this work would not be possible without the large gap in RTe$_3$. As shown in ref. 20, the evolution of the shadow bands can be followed from k$_x$=0.3a* up to 0.45a*, making their assignment unambiguous.

The real shape of the FS can be traced from the crossing position of these main and shadow bands. It is more complicated than in Fig. 17, because of the additional interactions between p$_x$ and p$_z$ and with the folded FS. The interaction between p$_x$ and p$_z$ is visible in Fig. 23a, it opens a gap of about 2V$_{px-pz}$≈0.3eV at their crossing, which replaces the linear dispersions near E$_f$ by a shallow parabola. This is in good agreement with the band calculation (see for example Fig. 2, where 2V$_{px-pz}$≈0.2eV). Taking this interaction into account, the FS evolves from the 2D TB FS of Fig. 4a to the square and outer sheets of Fig. 4b (for clarity, a large V$_{px-pz}$≈0.3eV is used in this figure). The 3D FS is obtained by folding the 2D with respect to the 3D BZ, as sketched on Fig. 4c. The main and folded bands also interact. This can be seen in Fig. 2b, where a gap 2V$_{3D}$≈0.18eV opens at their crossing, which takes place at –0.8eV. This is a similar strength as for the CDW interaction and, indeed, in Fig. 23, folded bands appear with similar intensity as the CDW shadow bands. After interaction (Fig. 24a), the outer part breaks into a small oval pocket near the zone boundary and a larger squared feature. The oval pocket is clearly present in the experimental data (see Fig. 24c). The periodicity is the one of the 3D BZ, but the distribution of the spectral weight is reminiscent of the 2D FS.[20] Once again, we take the size of the markers proportional to the spectral weight.

In Fig. 24, we proceed to the full reconstruction of the FS in the CDW state. This is similar to Fig. 17 but we now use the real FS of Fig. 24a instead of that of Fig. 4a. The main effect is a gapping of a large stripe along k$_z$ for about –0.25a*<k$_x$<0.25a*. The remaining fraction of the square is "closed" at the bottom by the shadow FS. The structure of the top of the square is more complicated. In the case of CeTe$_3$ (Fig. 24c), it clearly does not close but smoothly connects to the shadow FS. However, this shape sensitively depends on the relative strength of V$_{px-pz}$ and V$_q$, and also probably on the bilayer splitting. For YTe$_3$ (Fig. 5), the top part of the square is clearly closed. The fact that one can sort out these details is mainly the consequence of much improved resolution and data rate in modern ARPES. To close the other side of the pocket, along the outer part, the mechanism is similar to the one band case of Fig. 17.

One may wonder if the interaction at q$_{cdw}$ is equivalent to that at (c*-q$_{cdw}$), as we have seen that it is only when the weak folded FS is considered that q$_{cdw}$ becomes meaningful. If it is, we should observe in our data a gap at the crossing between the folded and CDW shadow bands. On the contrary, it seems in Fig. 23 that these two bands cross *without interacting*. In Fig. 24b, we calculate the FS in two extreme cases, with V(q$_{cdw}$)=0 (k$_x$>0) and V(q$_{cdw}$)= V(c*-q$_{cdw}$) (k$_x$<0). The crossing between folded and shadow FS is indicated by black arrows and this region is quite different in the two cases. The comparison with the experimental map of Fig. 24c clearly favors the first case.

This is a particularity of the CDW in this system, which is dominated by the in-plane coupling and is in fact essentially 2D. If not properly recognized, this could mimic a deviation from a sinusoidal distortion. In ref. 16, it was proposed that the CDW is commensurate within discommensurate domains, in order to create patterns of alternating short and long Te bonds. Such a picture would give more importance to the structural part of the CDW and implies deviations from the model of a nesting driven CDW. However, as shown in ref. 18, the different couplings at (c*-$q_{cdw}$) and $q_{cdw}$ more likely account for the anomalies observed in ref. 16. We confirm this view and conclude that the CDW in $RTe_3$ is truly incommensurate and nesting driven, albeit strongly 2D. We note that the CDW properties evolve in a non-monotonous way for the heavy rare-earths (Dy-Yb), which are notably characterized by two successive CDW transitions.[17] This also gives rise to quite different FS reconstruction, which will be the subject of a different paper.[40]

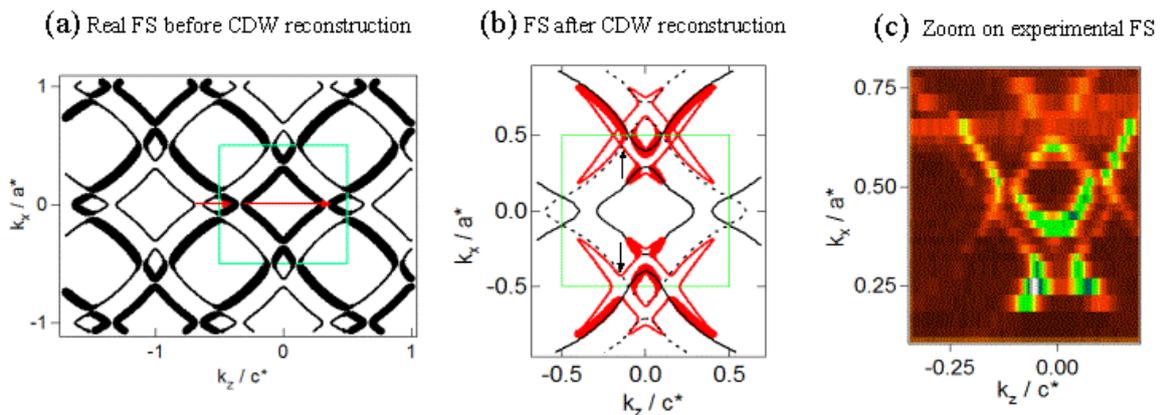

Fig. 24 : (a) 3D FS including interaction between p$_x$ and p$_z$ and between main and folded bands (bilayer splitting is omitted). The green square is the 3D BZ. The size of the markers is proportional to the spectral weight. The red arrows correspond to CDW wave vectors (c*-q$_{cdw}$) and q$_{cdw}$. (b) Red markers : weight of the reconstructed FS calculated within the TB model with V(q$_{cdw}$)=0 at k$_x$>0, and V(q$_{cdw}$)= V(c*-q$_{cdw}$) at k$_x$<0. Black contours are guide for the eyes of square and outer parts, dotted contours are for folded parts. (c) Zoom on the FS pockets measured in CeTe$_3$ with a photon energy of 55eV.

This study gives a precise topology of the FS that could be compared to the frequencies of quantum oscillations observed in these systems.[39] Another important consequence is the estimation of the fraction of ungapped FS. This is independent of the distribution of weight along the pockets, as, by definition, it is opposite on the pockets distant by $q_{cdw}$ (see Fig. 17). Therefore, it is sufficient to integrate the FS contour up to $k_x^0$ to obtain this fraction. We report on the right axis of Fig. 22 the percentage of the original FS contour that remains metallic. It increases from 35% in $CeTe_3$ to 55% in $DyTe_3$. This is not such a small fraction, which appears consistent with the good metallicity of the samples.[15,21] The linear term in the heat capacity was also found to be reduced to 37% in $LaTe_3$ and 60% in $YTe_3$, with respect to the calculated value, which is a similar order of magnitude.[21] On the other hand, a much smaller fraction (about 5%) was deduced from the analysis of optical measurements[22], a discrepancy that needs to be understood.

### F. Stabilization of the CDW

So far, we have given a full description of the CDW in $RTe_3$ as a function of the interaction parameter V. To close the loop, it would be desirable to see which strength of the electron-phonon coupling is required to stabilize such a CDW and whether the change of gap can be simply explained by the change of $n(E_f)$ due to lattice expansion.

In the traditional nesting scenario, the CDW is stabilized when the gain of electronic energy overcomes the loss of elastic energy. The gain of electronic energy in the CDW state can be easily calculated with the interacting TB model of III.A. It is due to the opening of the gap, which lowers the energy of these electrons on well nested parts of the FS. Examples of the density of states (DOS) are given in Fig. 25a, for one band, $t_{para}$=-1.8eV and $t_{perp}$=0.35eV, and different values of V. It has to be multiplied by 4 to obtain the DOS per cell (because of the 2 bands and 2 Te planes per cell). We take $q_{cdw}$ from formula of III.C for self consistency. As expected, a gap of maximum value $\Delta$=2V opens. $N(E_f)$ is reduced to 72% of the initial value at V=0.1eV and 55% at V=0.2eV. This is not directly comparable to our previous estimation of ungapped FS, as $n(E_f)$ is not constant along the FS, but confirms that a significant residual metallic character is expected. With such a model, we can then calculate the electronic energy as a function of V for any $t_{para}$ parameter, i.e. any value of $n(E_f)$.

To calculate the stable value of V, we subtract the elastic energy. It is a quadratic function of V :

$$E_{dis} = \frac{1}{2}M\omega_q^2\left(\langle x_q \rangle^2 + \langle x_{-q} \rangle^2\right) = \frac{n(E_f)}{2\lambda}V^2$$

where $\lambda$ is defined to be a dimensionless electron-phonon coupling constant $\lambda = g^2 n(E_f)/\hbar\omega_q$ ($g$ is defined in III.A and relates $\langle x_q \rangle$ to V). We find that $E_{dis}$=0.87*$V^2$ allows to find gap values in the range of those observed experimentally. Results are shown by red filled circles as a function of $t_{para}$ in Fig. 25b. The zone expected for $t_{para}$ is shaded, as well as the one corresponding to measured values of V deduced from Fig. 13. The theoretical variation is as large as the one observed experimentally, or even somewhat larger. This very large variation is due to the fact that, because of the imperfect nesting, increasing V does not only lower the energy of the electrons involved in the CDW but also increases their number. For V>0.3eV, the entire FS would be gapped and the increase of the gap value indeed slows down.

As the energies involved in this process are very small, this rough model can only be taken as indicative. The qualitative trend is however robust and it is worth checking the order of magnitude of the parameters we use. The value of $E_{dis}$ we use implies $\lambda$=0.23 (using the average TB value $n(E_f)$=0.4states/eV/Te plane), a moderately strong electron-phonon coupling, very similar to that observed in quasi-1D CDW systems, like $K_{0.3}MoO_3$ ($\lambda=0.34$) or $NbSe_3$ ($\lambda=0.53$).[1] The amplitude of the distortion in $CeTe_3$ was estimated to be $<x_q>=0.2$Å,[16] which is rather large, as compared for example to 0.05Å in $K_{0.3}MoO_3$ or 0.09Å in $(TaSe_4)_2I$.[1] As $\langle x_q \rangle = 1/\omega_q \sqrt{n(E_f)/2\lambda M} * V$, and V=0.2eV for $CeTe_3$ (ARPES value), we get $\omega_q \approx 5$meV with the parameters used so far and M, the mass of one Te atom. Although the values of frequencies involved in the CDW are not precisely known, this seems a reasonable order of magnitude and supports this model as the origin of the CDW in $RTe_3$.

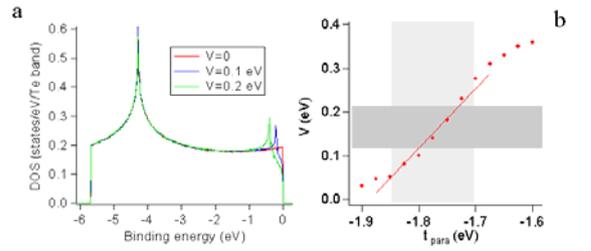

Fig. 25 : (a) Density of states in the TB model for indicated values of V ($t_{para}$=-1.8eV and $t_{perp}$=0.35eV). (b) Values of V stabilizing the CDW distortion as a function of $t_{para}$, assuming an elastic energy E=0.87*$V^2$. Shaded area are ranges sampled by $RTe_3$ compounds.

## IV. CONCLUSIONS

We have presented a detailed investigation of the electronic structure of $RTe_3$ based on a comparison between ARPES measurements and band calculations. The electronic structure is relatively simple, being built by broad p-type bands of Te atoms, where no strong correlation effects are expected. 3D couplings and the presence of rare-earths in the slabs separating the Te planes are shown to have only minor incidences on the electronic structure of the planes. Therefore, $RTe_3$ qualifies as a very interesting case study of a simple 2D metal for ARPES. The major complication is the occurrence of a bilayer splitting, due to the two planes per unit cell, which we analyse and compare to band calculations. Despite these advantages, we find that a detailed analysis of the ARPES lineshapes is difficult, probably because of an extrinsic broadening of the spectra.

On the other hand, $RTe_3$ definitely sets an ideal example of a nesting driven CDW system. We give a complete description of the ground state properties as a function of different R ions. We explain quantitatively the variation of the gap amplitude along the FS from its imperfect nesting, thanks to a simple TB model of the electronic structure. We show that the redistribution of the electronic density along the FS allows to maximize the gap value at the Fermi level up to $\Delta=2V$, where V is the electron-phonon interaction parameter. This implies a variation of the CDW wave vector $q_{cdw}$ with $\Delta$, which we calculate and find in very good agreement with the experimental variation. We detail the topology of the FS thanks to the observation of CDW shadow bands. We measure the extension of the residual metallic pockets as a function of the gap $\Delta$. The percentage of ungapped FS changes from 35% to 55%, depending on R.

Finally, we give a rough estimation of the balance between electronic and elastic energy needed to stabilize the CDW. We find that, for reasonable values of phonon frequencies ($\omega_q=5$meV) and electron-phonon coupling ($\lambda=0.23$), we can reproduce the variation of the gap with the density of states quite well. This certifies that such models are relevant for the description of these materials. It seems that the large gap of $RTe_3$ simply originates from the large displacements allowed in the square Te nets. The recent observation of the transition to the normal state around room temperature for many $RTe_3$ systems[17] and as a function of pressure in $CeTe_3$[23] opens interesting possibilities to fully characterize this state. The transition temperatures deviate notably from the mean-field expectation, as often observed in CDW systems,[1] which emphasizes the importance of fluctuations.